\documentstyle[preprint,aps,eqsecnum,epsf]{revtex}         
         
\begin{document}         
\title{{\bf {\huge Mean dynamical entropy of quantum system tends to infinity in         
the semiclassical limit}}}         
\author{{\ Wojciech S{\l}omczy{\'n}ski$^1$ and Karol \.Zyczkowski$^2$}}         
\address{\vspace{0.5cm} $^1$Instytut Matematyki,         
 $^2$Instytut Fizyki, Uniwersytet         
Jagiello\'nski, \\         
ul. Reymonta 4, PL-30 059, Krak\'ow, Poland}         
\maketitle         
         
\begin{abstract}         
We show that the mean dynamical entropy of          
a quantum map on the sphere tends         
logarithmically to infinity in the semiclassical limit.          
Consequences of this fact for classical dynamical systems          
are discussed.         
\end{abstract}         
         
\bigskip         
         
\vspace{0.5cm}         
         
\vspace{0.5cm}         
         
\begin{center}         
{\small e-mail: $^1$slomczyn@im.uj.edu.pl \quad $^2$karol@chaos.if.uj.edu.pl         
}         
\end{center}         
         
\vspace{0.5cm}         
         
\begin{center}         
{1991 MRSC: 81Q50, 28D20, 81R30, 58F11}         
         
{PACS: 05.45+b, 03.65.Bz}         
\end{center}         
         
\newpage \baselineskip=19pt         
         
\section{Introduction}         
         
Quantum analogues of classically chaotic systems have been an object of         
intensive investigations for almost twenty years. One has studied         
statistical properties of the spectra         
of quantized chaotic systems trying to prove         
that these systems can be described         
by suitable ensembles of random matrices         
(see e.g. \cite{G91,H91,CC96}).         
In this paper we follow the opposite         
direction: studying a generic quantum         
system we find strong arguments which         
support the conclusion that the         
dynamical entropy of the corresponding         
classical system is positive and,         
actually, arbitrary large. More precisely,         
we analyze the set of all {\sl structureless}          
(without geometric or time         
reversal symmetries) quantum systems \cite{LS97}.          
For these systems, described          
by the ensemble of unitary matrices, we compute the mean         
dynamical entropy averaged over the Haar measure. We show that it increases         
logarithmically with the dimension of the Hilbert space where a system         
lives, and so tends to the infinity in the semiclassical limit.         
         
The dynamical entropy we used depends         
on the choice of {\sl coherent states}.         
We prove our assertion for the families of $SU(d)$         
coherent states with $d\ge 2$. This includes         
 the well known spin $SU(2)$         
 coherent states, where         
the corresponding classical phase space is the two-dimensional sphere $S^2$.         
In the present work we only outline the main ideas of the proof putting         
apart the details of our reasoning to a forthcoming publication.         
         
\section{CS-entropy}         
         
The attempts to give a quantum analogue of the classical Kolmogorov--Sinai         
(KS) entropy have a rich history \cite{OP93,SZ94,R95}. Most of these         
definitions do not provide         
 a good framework for investigating the {\sl         
quantum chaos}. Some of them, as the Connes-Narnhofer-Thirring         
entropy \cite{CNT87} or the Alicki-Fannes \cite{AF94} entropy, vanish for         
finite quantum systems and can be rather applied in quantum statistical         
mechanics. Others do not give the correct semiclassical limit. Most of them         
are not easy to calculate neither analytically nor numerically, but for very         
simple cases. In a series of papers \cite{SZ94,KSZ97,S97} we proposed a new         
definition of dynamical quantum entropy based on the notion of coherent         
states, which we would like to recall now. Our approach to quantum entropy         
is based on the assumption that the knowledge of the time evolution of a         
quantum state is obtained by performing a sequence of approximate quantum         
measurements. The possible results of the measurement forms the         
(coarse-grained) stochastic phase-space being the quantum counterpart of the         
classical one. The evolution of the system between two subsequent         
measurements is governed by a unitary matrix.         
         
\smallskip Let ${\cal H}$ be an $N$-dimensional Hilbert space (which         
represents the {\sl kinematics} of the quantum system), let $U$ be a unitary         
operator on ${\cal H}$ (which describes the {\sl dynamics}{\it \/}), let $%
\Omega $ be a compact {\sl phase space}{\it \/} endowed with a probability         
measure $m$ (we shall write $dx$ for $dm(x)$), and let $\Omega \ni         
x\longrightarrow |x\rangle \in {\cal H}$ be a {\sl family of coherent states         
}(which are related to an {\sl approximated quantum measurement\/}), i.e., $%
\int_\Omega |x\rangle \langle x|\,dx=I$ and $x\longrightarrow |x\rangle $ is         
continuous. Moreover, we assume in this paper that $\langle x|x\rangle         
\equiv N$.         
         
Let ${\cal A}=\{E_1,\dots ,E_k\}$ be a partition of $\Omega $. We define the         
{\sl coherent states (CS) entropy of}{\it \ }$U$ {\it {\sl with respect to         
the partition} }${\cal A}$ by the formula         
\begin{equation}         
H(U,{\cal A}):=\lim_{n\to \infty }(H_{n+1}-H_n)=\lim_{n\to \infty }{\frac 1n}%
H_n\text{ ,}  \label{CSE1}         
\end{equation}         
where the partial entropies $H_n$ are given by         
\begin{equation}         
H_n:=\sum_{i_0,\dots ,i_{n-1}=1}^k-P^{CS}(i_0,\dots ,i_{n-1})\ln         
P^{CS}(i_0,\dots ,i_{n-1})\text{ ,}  \label{CSE2}         
\end{equation}         
for $n\in {\bf N}$; the probabilities $P^{CS}(i_0,\dots ,i_{n-1})$ of         
entering the cells $E_{i_0},...,E_{i_{n-1}}$ are         
\begin{equation}         
P^{CS}(i_0,\dots ,i_{n-1}):=\int\limits_{E_{i_0}}\,dx_0\dots         
\int\limits_{E_{i_{n-1}}}dx_{n-1}\prod_{u=1}^{n-1}K_U(x_{u-1},x_u)\text{ ,}         
\label{PRO1}         
\end{equation}         
for $i_j=1,\dots ,k,\ j=0,\dots ,n-1$; and the kernel $K_U$ is given by         
\begin{equation}         
K_U(x,y):={\frac 1N}|\langle y|U|x\rangle |^2  \label{KER1}         
\end{equation}         
for $x,y\in \Omega $. The kernel $K_U(x,y)$ may         
be interpreted as the $y$-dependent         
 {\sl Husimi distribution} of the transformed state $U|x\rangle $.         
If $U$ equals to the identity operator $I$, the quantity $K_I(x,y)$ is         
called the {\sl overlap} of coherent states $|x\rangle $ and $|y\rangle $.         
         
Note that both sequences in (\ref{CSE1}) are decreasing and the quantity $%
H_1=-\sum_{i=1}^km(E_i)\ln m(E_i)$, which does not depend on $U$, is just         
the {\sl entropy of the partition} ${\cal A}$. We denote it by $H({\cal A})$.         
         
There are two kinds of randomness in our model: the first is connected with         
the underlying unitary dynamics of the system; the second comes from the         
approximate measurement process. Accordingly, we divide CS-entropy with         
respect to a partition into two components: {\sl CS--measurement entropy}         
and {\sl CS--dynamical entropy}:         
         
\begin{equation}         
H_{meas}({\cal A}):=H(I,{\cal A})\text{ ;}{}  \label{MEAS}         
\end{equation}         
\begin{equation}         
H_{dyn}(U,{\cal A}):=H(U,{\cal A})-H_{meas}({\cal A})\text{ .}  \label{DYN1}         
\end{equation}         
Finally, we define the partition independent {\sl CS--dynamical entropy of}%
{\it \ $U$\/} as         
\begin{equation}         
H_{dyn}(U):=\sup_{{\cal A}}H_{dyn}(U,{\cal A})\text{ ,}  \label{DYN2}         
\end{equation}         
the supremum being taken over all finite partitions.         
         
It is conjectured that in the semiclassical         
limit the CS-dynamical entropy         
tends to the KS-entropy, if the unitary dynamics         
 comes from an appropriate         
quantization procedure (some results in this         
 direction were proved in \cite         
{SZ94}). In \cite{KSZ97,S97} we study         
 the properties of CS-dynamical entropy         
and present the methods of its numerical         
 computing based on the concept of         
iterated function systems (IFS).         
 In this work we evaluate the mean value of         
CS-dynamical entropy $\left\langle H_{dyn}(U)\right\rangle _{U(N)}$, taking         
the average over the unitary matrices $U(N)$ of the circular unitary         
ensemble (CUE).         
         
\section{SU($d$) - coherent states}         
         
We study CS-dynamical entropy for the family of         
 $SU(d)$  coherent states,         
 $d\geq 2$ \cite{GS93}.         
 Let $SU(d)\ni x\longrightarrow T_x\in U({\cal H}_M)$ be         
 the irreducible representation of the group $SU(d)$ in the group of         
unitary operators acting on Hilbert space ${\cal H}_M$, where $dim({\cal H}%
_M)=N=\bigl( {%
{M+d-1 \atop M}%
}\bigr)$, $M=1,2,...$. We can identify the phase space $\Omega $ with the         
coset space $SU(d)/U(d-1)$, where $U(d-1)$ is the maximal stability subgroup         
of $SU(d)$ with respect to the reference state $|\kappa \rangle \in {\cal H}%
_M$, i.e., the subgroup of all elements of $SU(d)$ which leave $|\kappa         
\rangle $ invariant up to a phase factor. The coherent states are defined by         
$|x\rangle =$ $T_x|\kappa \rangle $ for $x\in SU(d)$. The space $\Omega $         
which plays the role of the phase space of corresponding classical mechanics         
is isomorphic to the complex projective space $CP^{d-1}$. Hence, each point         
of $\Omega $ can be interpreted as a pure quantum state in a $d$ -         
dimensional complex Hilbert space ${\bf C}^d$. One can show that the overlap         
of two coherent states related to pure quantum states $\varphi $ and $\psi $         
is given by $|\langle \langle \varphi |\psi \rangle \rangle |^{2M}/N$, where         
$\langle \langle \cdot |\cdot \rangle \rangle $ is the canonical scalar         
product in ${\bf C}^d$ \cite{GS93}. The {\sl semiclassical limit} is         
obtained when $M\rightarrow \infty $, and $M^{-1}$ plays the role of the         
relative Planck constant. The above construction may be treated as a         
particular case of the general construction of group-theoretic coherent         
states \cite{P86}.         
         
If $d=2$, then $\Omega =SU(2)/U(1)$ is simply isomorphic to the         
two-dimensional sphere $S^2$, and the coherent states are ordinary {\sl spin         
coherent states} (see \cite{P86} and also \cite{KSZ97}). In this case $dim(%
{\cal H}_M)=N=M+1=2j+1$, where $j=\frac 12,1,\frac 32,...$ is the spin         
quantum number, the operators $T_x$ are represented by the Wigner rotation         
matrices, and for the state $|\kappa \rangle $ one usually takes the maximal         
eigenstate $|j,j\rangle $ of the component $J_z$         
of the angular momentum operator.         
         
\section{Continuous entropy}         
         
Computing the CS-dynamical entropy requires the time limit: $n\rightarrow         
\infty $. Surprisingly, one can obtain bounds for this quantity analyzing         
the {\sl continuous entropy} of $U$, which depends only on the one-step         
evolution of the quantum system:         
         
\begin{equation}         
H_U:=-\int\limits_\Omega \int\limits_\Omega {\sl K}_U(x,y)\ln {\sl K}%
_U(x,y)dxdy\text{ .}  \label{CONT1}         
\end{equation}         
         
This quantity is related to the ''classical-like'' entropy introduced to         
quantum mechanics by Wherl in \cite{W91}. Namely, $H_U$ is equal to the         
difference of the Wherl entropy of the states $U|x\rangle $ averaged over         
all $x$ from $\Omega $ and $\ln N$ (the latter term follows from the         
normalization in (\ref{KER1})). A similar quantities         
 have been also studied         
by Schroeck \cite{S85} (under the name of stochastic quantum         
mechanical entropy) and by Mirbach and Korsch \cite{MK95}. Calculation of         
continuous entropy is particularly easy for $U=I$ and $SU(d)$ coherent         
states. In this case ${\sl K}_I(x,y)=         
|\langle \kappa | T_y^{-1} T_x|\kappa \rangle |^2 /N$         
 for any points $x,y$ belonging to         
  the phase space $\Omega $ and so         
         
\begin{equation}         
H_I=-\int_\Omega \int_\Omega \frac{|\langle \kappa |T_{y^{-1}x}|\kappa         
\rangle |^2}N\ln \frac{|\langle \kappa |T_{y^{-1}x}|\kappa \rangle |^2}N%
dx\,\smallskip dy=-\int\limits_\Omega \frac{|\langle \kappa |T_z|\kappa         
\rangle |^2}N\ln \frac{|\langle \kappa         
|T_z|\kappa \rangle |^2}Ndz .         
\label{MEAS2}         
\end{equation}         
         
We can now apply the formula for the overlap of two $SU(d)$ coherent states         
(see above), which enables us to use the result from Jones \cite{J90,J91}         
who calculated generalized mean entropy of pure quantum state in a $d$%
-dimensional complex Hilbert space. Proceeding in this way we get         
         
\begin{equation}         
H_I=-\ln N+M\left[ \Psi \left( M+d\right) -\Psi \left( M+1\right) \right]         
\text{ ,}  \label{MEAS3}         
\end{equation}         
where $N=dim({\cal H}_M)=\bigl( {{M+d-1 \atop M} }\bigr)$         
 and $\Psi $ is the digamma function, satisfying $\Psi (x+1)=\Psi         
(x)+\frac 1x$ for $x>0$.         
If $d=2$ the above formula reduces to         
         
\begin{equation}         
H_I=-\ln N+\frac{N-1}N\text{ .}  \label{MEAS4}         
\end{equation}         
         
\section{Bounds for CS-dynamical entropy}         
         
Let ${\cal A}$ be a partition of $\Omega $ and let $U$ be a unitary operator         
on ${\cal H}_M$. Using classical methods from the information theory (see         
\cite{G77}, Sect. 2.2) we can prove the following observation:         
         
\begin{equation}         
\inf_{{\cal A}}\left[ H_{n+1}\left( U,{\cal A}\right) -H_n\left( U,{\cal A}%
\right) -H\left( {\cal A}\right) \right] =H_U  \label{BOUND1}         
\end{equation}         
for each natural $n$, where the coherent states partial entropies $H_n\left(         
U,{\cal A}\right) $ are defined in (\ref{CSE2}). Hence and from the         
definition of CS-entropy we get         
         
\begin{equation}         
H_U+H\left( {\cal A}\right) \leq H\left( U,{\cal A}\right) \leq H\left(         
{\cal A}\right)  \label{BOUND2}         
\end{equation}         
and         
\begin{equation}         
\inf_{{\cal A}}\left[ H\left( U,{\cal A}\right) -H\left( {\cal A}\right)         
\right] =H_U\text{ .}  \label{BOUND3}         
\end{equation}         
In fact, the infimum in (\ref{BOUND1})       
and (\ref{BOUND3}) is achieved if the maximal diameter of a member       
of the partition ${\cal A}$ tends to zero.       
Thus for a sufficiently fine partition the       
CS-entropy splits approximately into two parts: the one which       
depends only on the partition, and the other depending only on the dynamics.       
Combining the above formulae        
with the analogous obtained for $U=I$ we conclude that        
\begin{equation}         
-H_I+H_U\leq H_{dyn}(U)\leq -H_I\text{ .}  \label{BOUND4}         
\end{equation}         
         
The famous Lieb conjecture says that for $d=2$ the Wherl entropy attains its         
minimum equal to $H_I+\ln N=(N-1)/N$ (compare \ref{MEAS4}) for any       
coherent state (see         
\cite{L78}, and \cite{L88} for partial results). We conjecture that this is         
also true for $d>2$ and the minimum of the Wehrl entropy for $SU(d)$ is         
equal to $H_I$ given by (\ref{MEAS3}) plus $\ln N$.         
 This would imply $H_I\leq H_U$,         
and consequently $H_{dyn}(U)$ $\geq 0$ for every unitary matrix $U$.         
         
As we can see above the quantity $H_I$ decreases approximately as $-\ln N$         
and so, if the generalized Lieb conjecture is true, then the entropy $         
H\left( U,{\cal A}\right) $ is limited from below by $H\left( {\cal A}         
\right) -\ln N$. This agrees with the bound obtained by Halliwell for the         
information of phase space distributions derived from the probabilities for         
quantum histories \cite{H93}. Note, however,         
 that the bound (\ref{BOUND2})         
seems to be more precise, because, as we will show,         
 $-H_U$ is typically much smaller then $-H_I$.         
         
Averaging (\ref{BOUND4}) over the set of all unitary matrices $U(N)$ with         
respect to the Haar measure $\mu $ we get         
\begin{equation}         
-H_I+\left\langle H_U\right\rangle _{U(N)}\leq \left\langle         
H_{dyn}(U)\right\rangle _{U(N)}\leq -H_I\text{ ,}  \label{BOUND5}         
\end{equation}         
Thus, to obtain the desired bounds for the mean CS-dynamical entropy, it         
suffices to calculate $\left\langle H_U\right\rangle _{U(N)}$. We have         
         
\begin{equation}         
\left\langle H_U\right\rangle _{U(N)}=-\int\limits_{U(N)}\left(~         
\int\limits_\Omega \int\limits_\Omega K_U(x,y)\ln K_U(x,y)dxdy\right) d\mu         
(U)\text{ .}  \label{FINAL1}         
\end{equation}         
Since $K_U(x,y)=|\langle y|U|x\rangle |^2/N=|\langle \kappa         
|T_y^{-1}UT_x|\kappa \rangle |^2/N$, interchanging the order of         
integration and using the invariance of the Haar measure on $U(N)$       
we conclude that         
         
\begin{eqnarray}         
\left\langle H_U\right\rangle _{U(N)} &=&-\int\limits_\Omega         
\int\limits_\Omega \left( \,\int\limits_{U(N)}\frac{|\langle \kappa         
|T_y^{-1}UT_x|\kappa \rangle |^2}N\ln \frac{|\langle \kappa         
|T_y^{-1}UT_x|\kappa \rangle |^2}Nd\mu (U)\right) dxdy  \nonumber \\         
&=&-\int\limits_{U(N)}\frac{|\langle \kappa |V|\kappa \rangle |^2}N\ln \frac{%
|\langle \kappa |V|\kappa \rangle |^2}Nd\mu (V)\text{ .}  \label{FINAL2}         
\end{eqnarray}         
         
We can calculate the last quantity utilizing the formula for the         
distribution of $\langle \kappa |U|\kappa \rangle $ given by Ku\'{s} et al.         
\cite{KMH88}. Otherwise, we can use the already mentioned result of Jones         
\cite{J91}. Applying one of these methods we get the following formula:         
         
\begin{equation}         
\left\langle H_U\right\rangle _{U(N)}=-\ln N+\Psi \left( N+1\right) -\Psi         
\left( 2\right) \text{ .}  \label{FINAL3}         
\end{equation}         
         
Finally from (\ref{MEAS3}), (\ref{BOUND5}) and (\ref{FINAL3}) we obtain the         
main result of this work: a lower and an upper bound for the mean dynamical         
entropy         
         
\begin{equation}         
l_b~ \leq~ \left\langle H_{dyn}^{SU(d)}\right\rangle ~ \leq ~ u_b ,         
\label{BOUB1}         
\end{equation}         
where         
         
\begin{eqnarray}         
l_b &=& \Psi \left( N+1\right) -\Psi \left( 2\right)         
-M\left[ \Psi \left( M+d\right) -\Psi \left( M+1\right) \right],         
 \nonumber \\         
u_b &=&\ln N         
-M\left[ \Psi \left( M+d\right) -\Psi \left( M+1\right) \right],         
  \label{BOUB2}         
\end{eqnarray}         
with $N=\bigl( {%
{M+d-1 \atop M}%
}\bigr)$.         
         
In the semiclassical limit $M\rightarrow \infty $ we get simple         
approximations for both bounds         
         
\begin{equation}         
l_b\sim \ln N-d+\gamma \text{ ,}{\text{\quad }}\text{{and}}{\quad }u_b\sim         
\ln N-d+1\text{ ,}  \label{FINAL4}         
\end{equation}         
where $\gamma $ is the Euler constant.         
The difference between an upper bound (which is actually the         
maximal value of the CS-dynamical entropy!) and a lower one converges to the         
constant $1-\gamma ~ \simeq ~ 0.\,42278$ if $M\rightarrow \infty $. Hence the         
mean value of CS-dynamical entropy tends in the semiclassical limit to the         
infinity exactly as $\ln N$.         
         
Let us consider the case $d=2$, where the family of spin coherent states is         
parametrized by the points lying on the two-dimensional sphere $S^2$. The         
mean entropy of quantum maps on the sphere is thus bounded by         
         
\begin{equation}         
\Psi \left( N+1\right) -\Psi \left( 2\right) -1+ \frac{1}N \leq \left\langle         
H_{dyn}^{SU(2)}\right\rangle _{U(N)}\leq \ln N-1 + \frac{1}N.  \label{FINAL7}         
\end{equation}         
         
 \begin{figure}         
\unitlength 1cm         
\begin{picture}(9,8)         
\put(0.7,10.8){\includegraphics{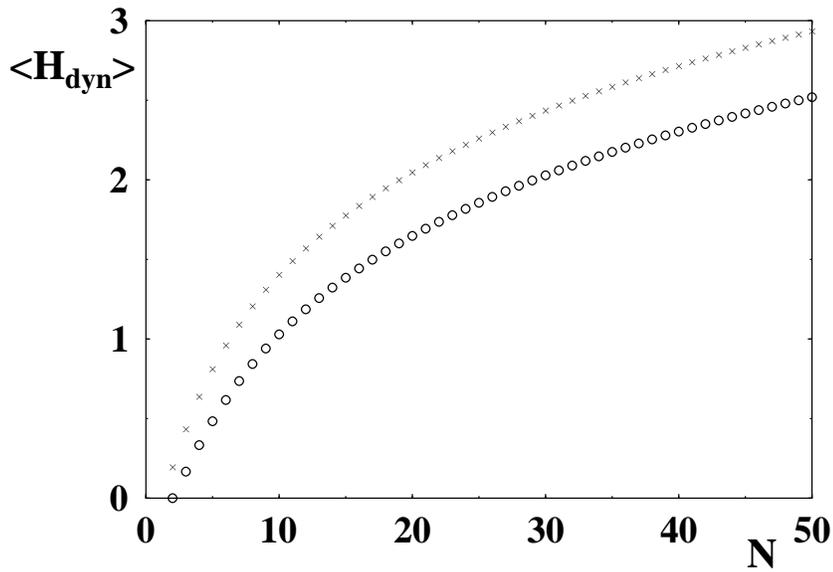}}         
 \end{picture}         
\unitlength 1bp         
\caption{Upper $(\times )$ and lower ($\circ$)         
 bounds for the mean CS-dynamical entropy of         
 unitary matrices representing structureless quantum         
systems on the sphere as a function of the matrix dimension $N=2j+1$. }         
\end{figure}         
\baselineskip=19pt         
         
The dependence of both bounds on the quantum number $N=2j+1$ is presented in         
Fig.~1. In the semiclassical limit $N\to \infty $ the mean dynamical entropy         
diverges in contrast to the CS-dynamical entropy of a given quantum map,         
which converges to the KS-entropy of the corresponding classical system.         
Therefore, for sufficiently large $N$ a matrix $F$ representing a given         
quantum map must differ from a generic (with respect to the Haar measure on $%
U(N)$) unitary matrix. To visualize the difference we present         
 in Fig.~2 the         
Husimi function of an exemplary         
 coherent state $|\vartheta ,\phi \rangle         
=|1.6,3.4\rangle $ transformed once by a Floquet operator         
 $F$ representing the         
kicked top \cite{H91} in the classically chaotic regime (a), and by a random         
unitary matrix $U$ (b). The sphere is represented in the Mercator projection         
with $0\le \phi <2\pi $ and $0\le \vartheta <\pi $, $t=\cos \vartheta $. In         
the former case the wave packet remains localized in the vicinity of the         
classical trajectory, while in the latter, it is entirely delocalized         
already after one iteration. The same data plotted in the log scale allow         
one to detect zeros of the Husimi functions \cite{LV90,WK97}. For the      
quantum map F they form a regular spiral-like structure (c) in contrast      
to the random distribution over the entire phase space for the unitary      
map $U$ (d).         
         
\begin{figure}         
\unitlength 1cm         
\begin{picture}(9.5, 9.5)         
\put(0.0,-4.2){\includegraphics{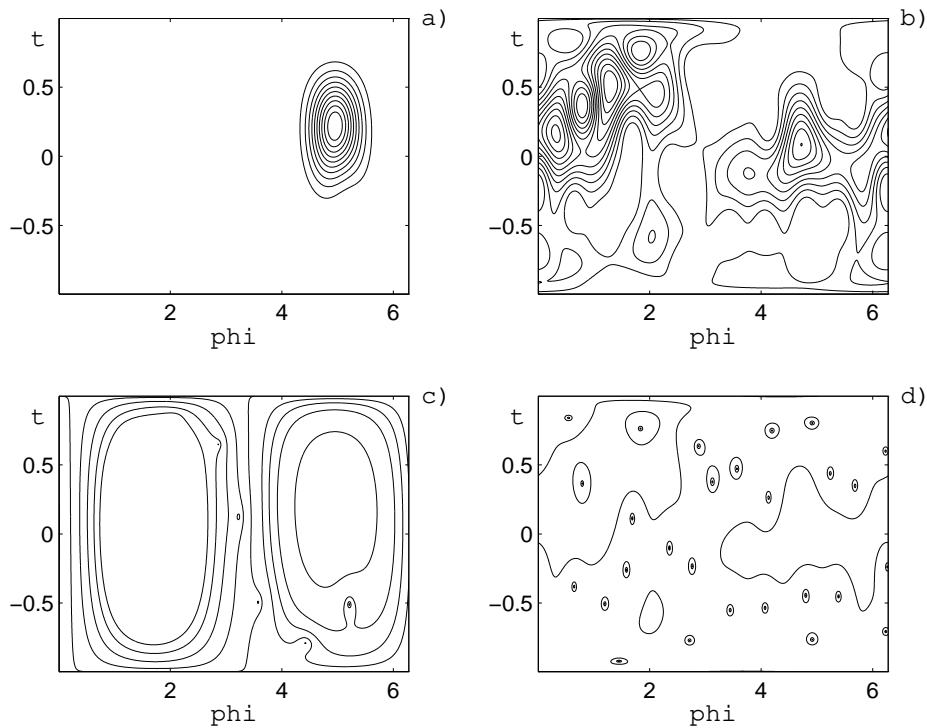}}         
 \end{picture}         
\unitlength 1bp         
\baselineskip=16pt         
\caption{Contour plot of the Husimi function of an exemplary coherent state         
transformed by the quantum kicked top map (a) and by a generic random         
matrix (b) for $N=30$. Observe qualitative differences in the         
distribution of zeros of the Husimi function visible in figures         
(c) and (d), respectively,         
obtained from the same data using a log scale for the contour heights. }         
\end{figure}         
         
\baselineskip=19pt         
         
\section{Conclusions}         
         
The estimate (\ref{BOUB1}) allows us to conclude that a quantum         
system represented by a typical unitary matrix from CUE ensemble is         
characterized by positive dynamical entropy, which is only insignificantly         
smaller than the maximal diverging with $M\sim 1/\hbar $. In other words,         
a generic quantum system is almost as chaotic, as possible. We prove this         
for $SU(d)$ coherent states, but the method seems to work also in the         
general case, i.e., for coherent states defined on arbitrary homogenous         
compact manifold, as well as for the orthogonal and symplectic ensembles.         
         
At a first glance this result seems         
 to be paradoxical as the KS-entropy of a         
classical map is finite and         
 the CS-dynamical entropy of the corresponding         
quantum system tends to this value in the semiclassical limit. Hence for a         
Hilbert space of sufficiently large dimension matrices representing a         
quantum analogue of a given classical chaotic system can not be typical.         
Their entropy is substantially smaller than the CUE average, even though         
many other statistics (level spacing distribution, spectral rigidity, etc.)         
conform to the predictions of random matrix theory.         
         
However, this need not contradict the general believe that quantum         
analogues of classically chaotic systems might be represented by a typical         
unitary matrix. The paradox can be resolved if we assume that strongly         
chaotic systems dominate less chaotic ones in the         
'space' of classical systems         
defined on the corresponding symplectic         
 manifold. Thus, our results provide         
a strong argument in favor of ubiquity of         
 chaos in the classical mechanics.         
         
\vskip 1.0cm         
{\sl Acknowledgments.}         
This work was supported by the Polish KBN grant P03B 060 13.         
         

\end{document}